\documentclass[prb,twocolumn,showpacs,preprintnumbers,amsmath,amssymb]{revtex4}

\usepackage{graphicx}
\usepackage{dcolumn}
\usepackage{bm}
\usepackage{subfigure}

\begin{document}


\title{Electronic transport properties through thiophenes on switchable domains}

\author{T. Kunze$^1$, S. Gemming$^1$, V. Pankoke$^1$, K. Morawetz$^{2,3}$, R. Luschtinetz$^4$, G. Seifert$^4$}
\affiliation{$^1$Forschungszentrum Dresden-Rossendorf, P.O. Box 51 01 19, D-01314, Germany}%
\affiliation{$^2$University of Applied Science M\"unster, Stegerwaldstrasse 39, D-48565 Steinfurt, Germany}%
\affiliation{$^3$International Center of Condensed Matter Physics, University of Brasilia, 70904-970, Brasilia-DF, Brazil}
\affiliation{$^4$Physical Chemistry, TU Dresden, D-01062 Dresden, Germany}%

\date{\today}

\begin{abstract}
The electronic transport of electrons and holes through stacks of $\alpha$,$\omega$-dicyano-$\beta$,$\beta$'-dibutyl- quaterthiophene (DCNDBQT) as part of a novel organic ferroic field-effect transistor (OFFET) is investigated. The novel application of a ferroelectric instead of a dielectric substrate provides the possibility to switch bit-wise the ferroelectric domains and to employ the polarization of these domains as a gate field in an organic semiconductor. A device containing very thin DCNDBQT films of around 20 nm thickness is intended to be suitable for logical as well as optical applications. We investigate the device properties with the help of a phenomenological model called multilayer organic light-emitting diodes (MOLED), which was extended to transverse fields. The results showed, that space charge and image charge effects play a crucial role in these organic devices.
\end{abstract}

\maketitle

\section{\label{sec:Introduction}Introduction}

The investigation of organic semiconductors, which consist of special organic molecules or polymers, has become more and more attractive in recent years \cite{Dim02,Fac07}. From the experimental point of view, the main focus was directed to the fabrication and characterization of suitable organic devices such as organic light-emitting diodes (OLEDs) and organic field-effect transistors (OFETs) \cite{Yas05,Koy06,Hau08}. 
Theoretical effort was linked to the description of the physical background, which includes the investigation of the underlying transport mechanisms. A mobile charge carrier can be transported ballistically along a polymer or it jumps between suitably stacked molecules, which is well-known as hopping transport \cite{Bae93,Dun96,Nov98,Her08,Mor08,Mor09,Fis08}. In addition, the characterization of the organic-electrode interface is essential for potential applications \cite{Har95,Nos06}.

\begin{figure}[ht]
        \centering
	\includegraphics[width=8.3cm]{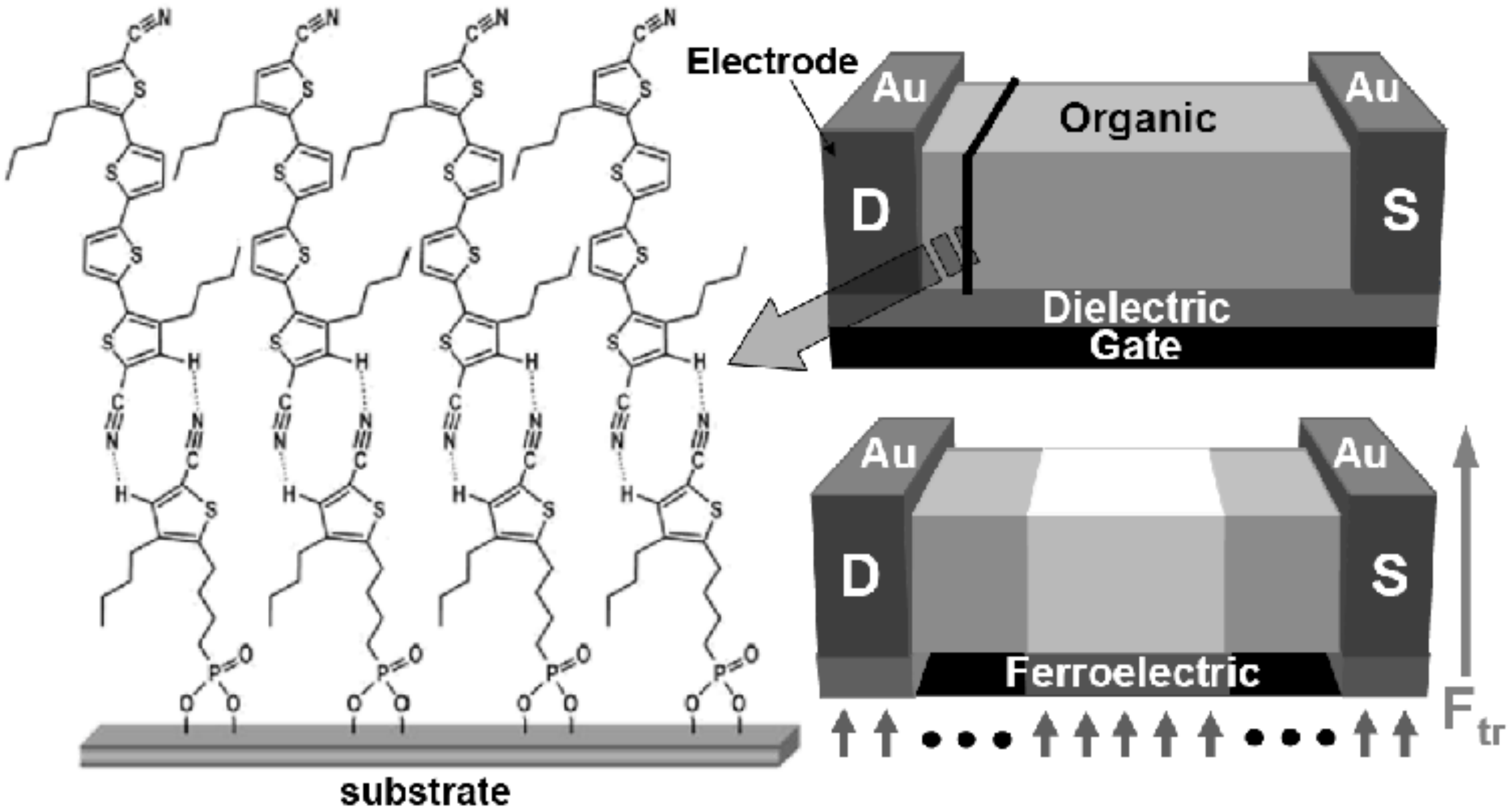}
        \caption{\label{fig:generalDevice}[left] Enlarged slice of the thiophene film representing its morphology \cite{Hau08}. [right] General setup of an OFET device (upper panel) and the modified version (lower panel), where the dielectric substrate is replaced by a ferroelectric material such as PZT. The ferroelectric domain induces a polarization field which influences the organic material. Grey and black arrows at the bottom represent different directions of polarization.}
\end{figure}

We present theoretical investigations of a novel organic device, which can be formally denoted as an organic ferroic field effect transistor (OFFET). Contrary to the dielectric substrates generally employed in conventional OFET setups (see upper panel of figure \ref{fig:generalDevice}), OFFET devices are fabricated with ferroelectric materials such as BaTiO$_3$ or Pb(Zr,Ti)O$_3$. Below a material-dependent critical temperature $T_c$, the ferroelectric material spontaneously polarizes and forms ferroelectric domains with different directions of polarization. In the OFFET this induced polarization field serves as gate field. Due to the different operation principle, this can be seen as an alternative to the well-established high performance ferroelectric field-effect transistor (FeFET) devices \cite{Nab05,Nab06,Yil07,Ngu07,Laz09,Tie09}. 

The ideal ferroelectric substrate would form domains with two preferable directions of polarizibility. Nowadays the orientation of these ferroelectric domains can be resolved for instance with the help of near-field microscopy \cite{Keh08}. The ideal ferroelectric substrate would form domains with two preferable directions of polarizability, in-plane and orthonormal. An in-plane polarization is expected to have only minor influence on the organic medium, whereas an polarization field $\vec{F}_{tr}$ in orthonormal direction is supposed to have maximum influence. Consequently, the organic material must be polarizable in this orthonormal direction. Therefore, it is essential for potential applications to understand the influence of the ferroelectric domains on the organic material. Thus, we will study the polarizability of the organic medium.

The employed organic material will be $\alpha$,$\omega$-dicyano-$\beta$,$\beta$'-dibutylquaterthiophene (DCNDBQT). This aromatic molecule, belonging to the group of the oligothiophenes, can form well-ordered stacks with a good $\pi$-orbital overlap. This high degree of ordering can favor high carrier mobilities and additionally decrease disorder effects, which are typically observed in disordered organic media \cite{Ver03}. The DCNDBQT films show thermal stability after preparation with an energy gap $E_g$ in the range of semiconducting materials ($E_g < 3 eV$) \cite{Hau08}.

Furthermore, we will present the investigation of the electronic transport in such an OFFET device. For this purpose, the device is simulated with the help of a phenomenological numerical model called multilayer organic light-emitting diode (MOLED) \cite{Hou03,Ber05} and its extension to transverse fields. While transport properties can be found elsewhere \cite{Mor09}, we will now focus on electric properties influenced by the ferroelectric gate. In addition, the simulation-relevant parameters are determined either by experimental, analytical or computational methods.

Another crucial point regards the carrier transport in organic systems, which can be very complex due to dynamic polarization effects, either in the organic medium itself and additionally in the dielectric substrate \cite{Hou06}. Due to the low carrier densities in our system, these intrinsic polarization effects will play only a minor role. Additional experiments are necessary to clearify this assumption.



Generally, the magnitude of the gate voltage is constant and depends on the employed ferroelectric material. Thus, the gate voltage in our simulations is a simulation parameter, which has later to be identified with the intrinsic, material-dependent surface polarization. The material-intrinsic electric field can be roughly estimated by employing $\vec{F}_{tr}=\frac{\sigma}{2\epsilon_0 \epsilon_r}$ under the assumption of a homogeneously charged ferroelectric surface. Here, $\sigma$ represents the surface charge, which lies typically in the range of $30 \frac{\mu C}{cm^2}$ for materials like Pb(Ti,Zr)O$_3$ and LiNbO$_3$, while $\epsilon_r=2.8$ denotes the relative dielectric constant of the molecular crystal of thiophene \cite{Ost28}. The estimated polarization field amounts to $\vec{F}_{tr} \approx 6 \frac{GV}{m}$ directly at the interface between substrate and organic. Furthermore, a decrease of this electric field is expected further inside the organic medium due to screening effects. This might limit the organic film thickness, particularly in logical devices, where switching between two states is important.

Ferroganic devices have a great potential for optical as well as logical applications. While optical ferroganic devices are interesting due to the possibility of band-gap engineering (see chapter 2 for detailed explanation), OFFET structures take benefit of the domain stability in a working device. This advantage is given by the material-specific gate field which persists even after a power outage.

Section \ref{Molecule polarization} presents the influence of the polarization field on the molecule. Section \ref{Numerical details} gives some information about the alorithm employed in the MOLED model and the determination of the simulation parameters. Sections \ref{Results} and \ref{Conclusion} show the results of the simulation and a conclusion, respectively.

\section{Molecule polarization}
\label{Molecule polarization}

The modeling of the OFFET requires a detailed understanding of how the molecule DCNDBQT will be influenced by the application of a transverse electric field $\vec{F}_{tr}$, which is generated by the ferroelectric material. Therefore, the field-induced variation of the highest occupied molecular orbital (HOMO) and the lowest unoccupied molecular orbital (LUMO) is of particular interest because these orbitals correspond to the valence and conduction band of inorganic semiconductors, respectively.

\begin{figure}[ht]
	\centering
	\subfigure[]{\label{fig:HOMO_LUMO_variation_a}\includegraphics[height=2.2cm, angle=90]{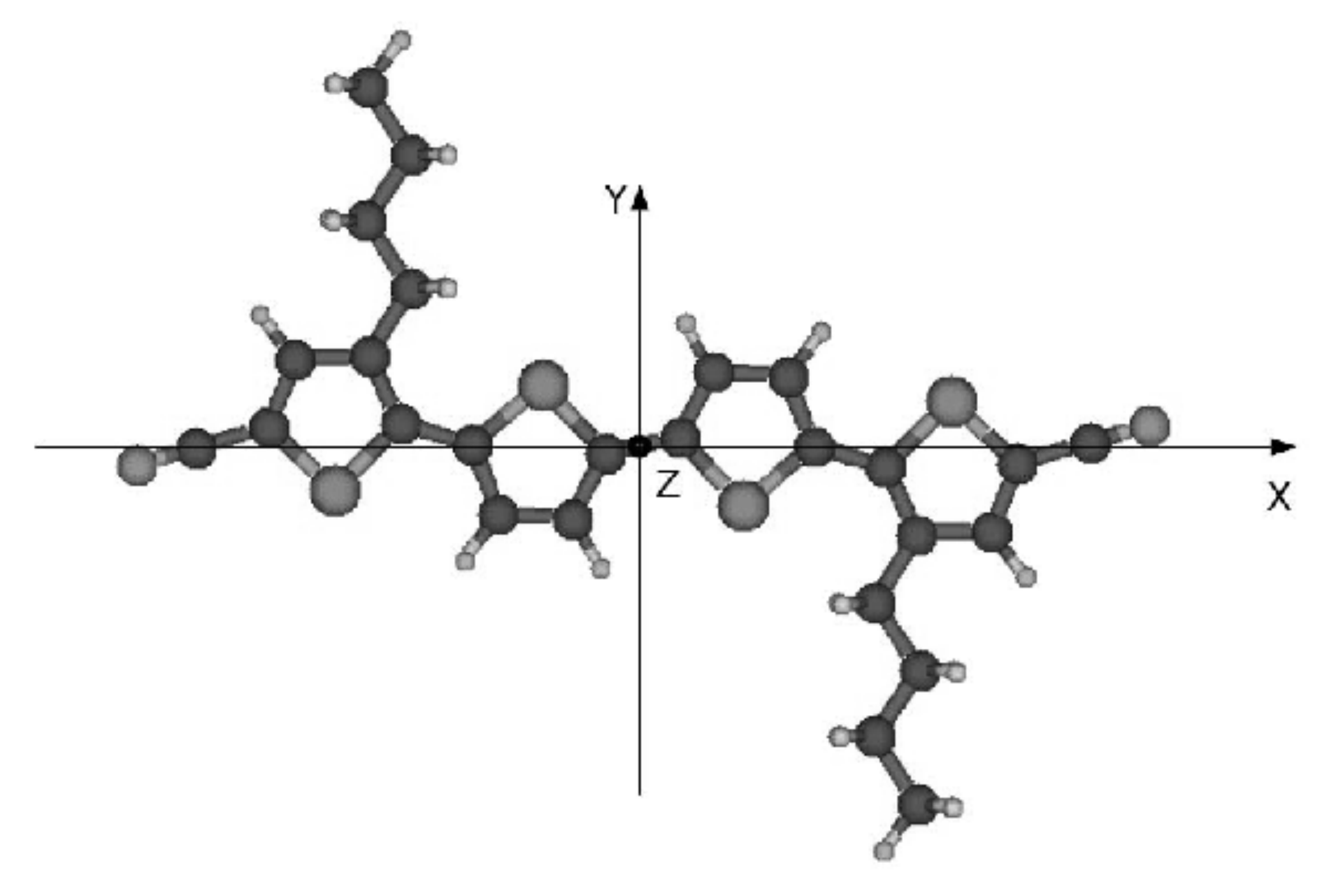}}
	\subfigure[]{\label{fig:HOMO_LUMO_variation_b}\includegraphics[height=3.4cm]{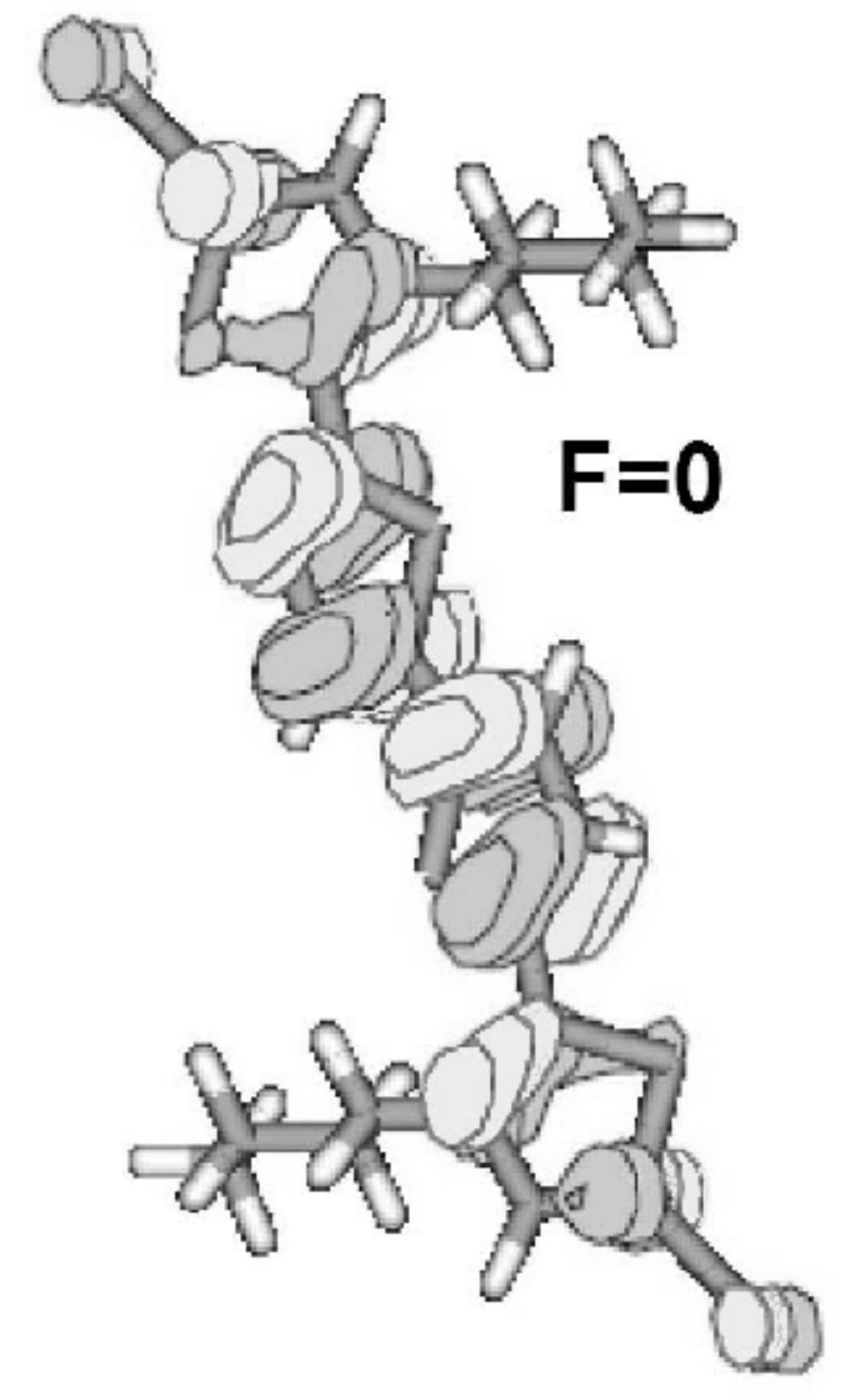}}
	\subfigure[]{\label{fig:HOMO_LUMO_variation_c}\includegraphics[height=3.4cm]{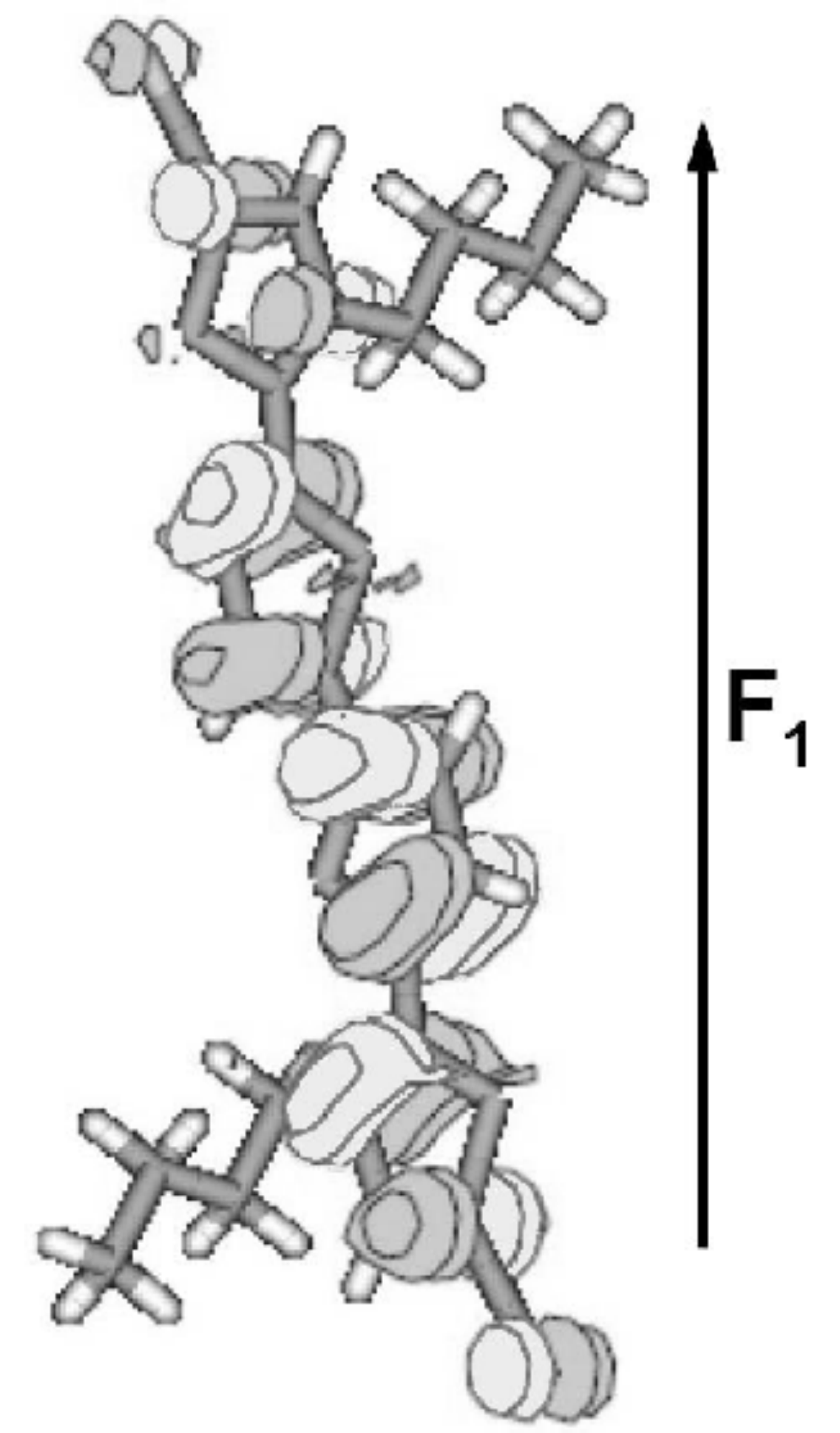}}
	\subfigure[]{\label{fig:HOMO_LUMO_variation_d}\includegraphics[height=3.4cm]{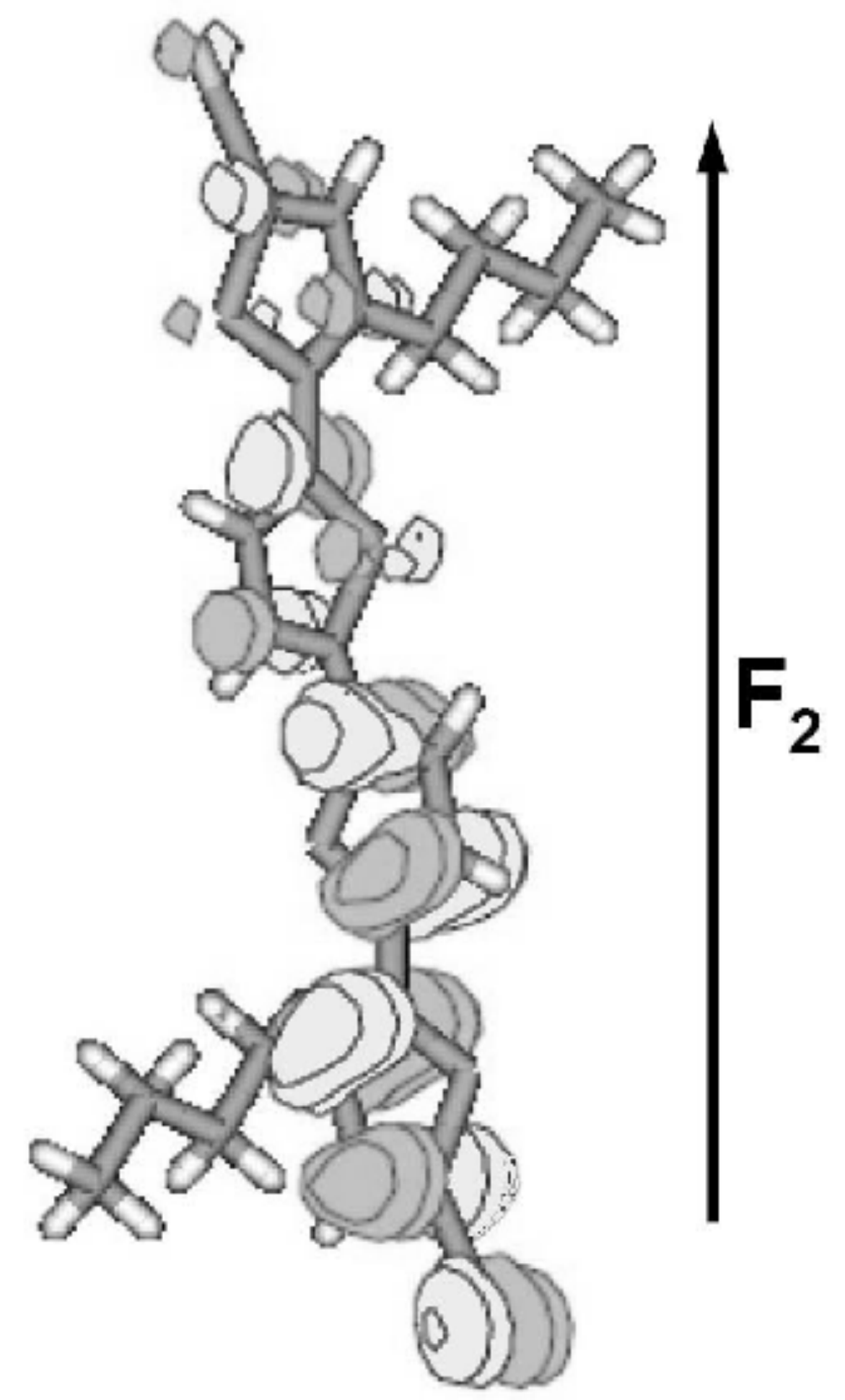}}
	\caption[DCNDBQT HOMO deformation under presence of an applied electric field]{\label{fig:HOMO_LUMO_variation}DFTB-calculated HOMO deformation of DCNDBQT by the influence of a transverse electric field in positive x direction. The different colors denotes the different signs (+/-) of the electron wave function. (a): DCNDBQT with coordinate system. (b)-(d): Increase of the field strength from zero field, over $F_1$ = 2 $\frac{GV}{m}$ up to $F_2$ = 6 $\frac{GV}{m}$.}
\end{figure}

\begin{figure}[ht]
        \centering
        \includegraphics[width=8cm]{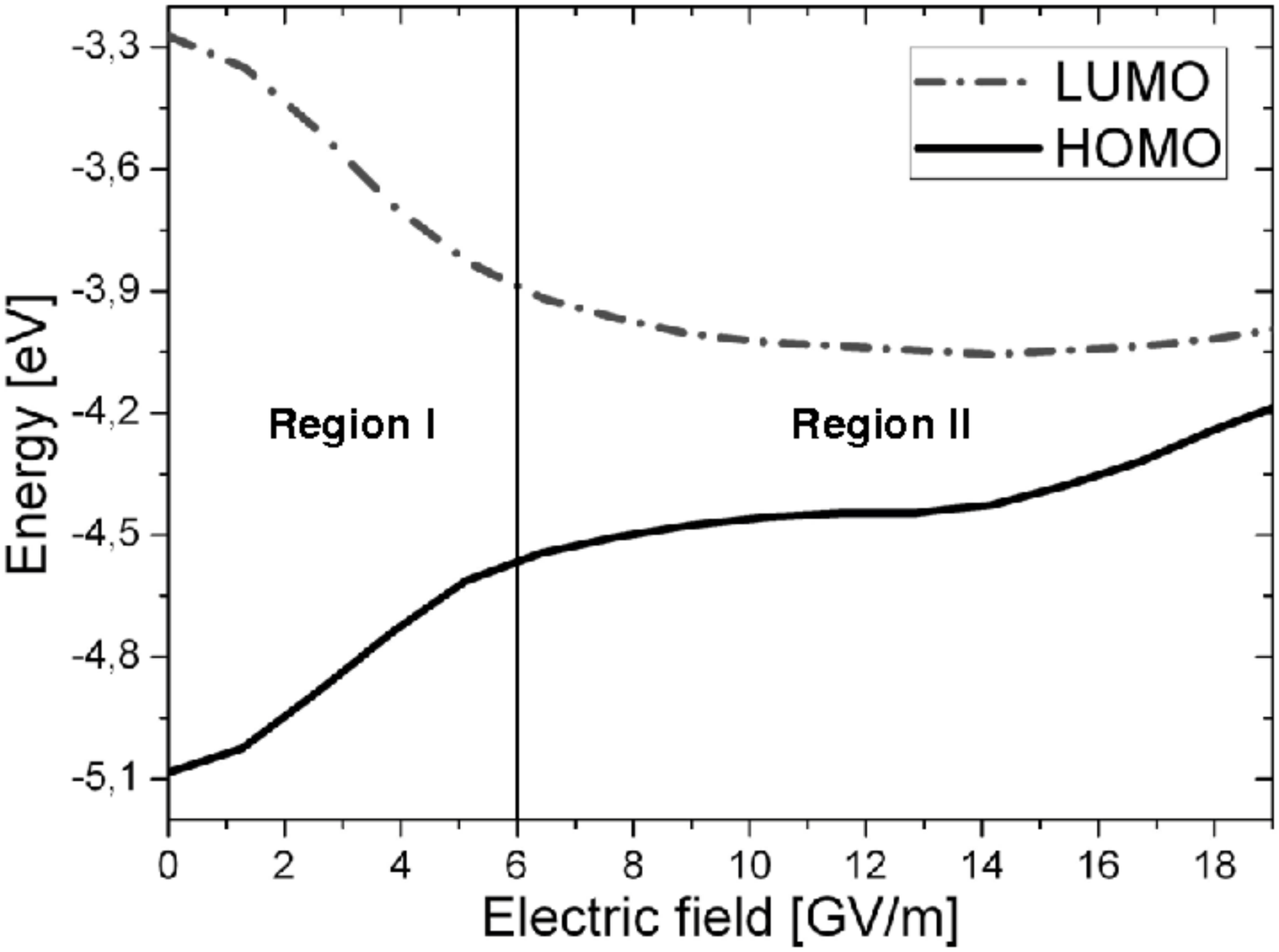}
        \caption{\label{fig:HOMO_LUMO_varDiagram} Variation of the HOMO/LUMO energy levels of quaterthiophene under the application of a transverse field $\vec{F}_{tr}$. The band gap decreases linearly (region I) with increasing strength of the electric field. The non-linear regime (region II) is reached at fields higher than 6 $\frac{GV}{m}$, when the HOMO-LUMO compositions change.}
\end{figure}

For this purpose, the influence of the transverse electric field on the electronic properties of oligothiophenes has been studied using the self-consistent charge density-functional based tight binding method SCC-DFTB \cite{Fra02,Gem07a,Gem07b}. Figure \ref{fig:HOMO_LUMO_variation} illustrates, that the $\pi$-electron system is most strongly deformed along the molecule with increasing $\vec{F}_{tr}||x$, hence polarizing the molecule towards one short edge.

Figure \ref{fig:HOMO_LUMO_varDiagram} shows the resulting variation of the HOMO and LUMO energy levels of DCNDBQT under an applied transverse electric field $\vec{F}_{tr}||x$. This field $\vec{F}_{tr}$ induces a compensating polarization in the molecule, which corresponds to a mixing of HOMO and LUMO levels. This leads to a linear decrease of the undistorted HOMO-LUMO gap, in detail from 1.8 eV at zero field to 0.6 eV at a high external field of 6 $\frac{GV}{m}$ (region I). Consequently, this molecule is suitable for devices which take advantage of a band-gap engineering.

Figure \ref{fig:HOMO_LUMO_varDiagram} also shows a transition to non-linear behavior at fields higher than 6 $\frac{GV}{m}$ (region II). For our simulations, we will stay in the linear regime by employing fields not higher than $\vec{F}_{tr}=4\frac{GV}{m}$.

\section{Numerical details}
\label{Numerical details}

We employ the phenomenological model MOLED, introduced by Houili et al. \cite{Hou03}, to investigate the charge carrier profile in our OFFET device. A main advantage of this model is its flexibility due to the full parametrization of the device. These parameters comprise, for instance, the electrode work functions, molecular energy levels, temperature and applied electric fields. 

In this section, we will present first the general concepts and boundary conditions of MOLED and second, the determination of the most important simulation parameters.

\subsection{General concepts and boundary conditions of MOLED}

MOLED is based on an effective 1-dimensional model, which can be applied, if the system fulfills three key assumptions. (i) The electrodes are planar and parallel to each other. (ii) The organic channel material between the electrodes is homogeneously distributed and parallel to the electrodes. (iii) The distance between the electrodes should be small compared to the other two spatial dimensions.

Experimentally, the first point can be achieved by photolithographic fabrication of the electrodes. Furthermore, the AFM image in ref. \onlinecite{Hau08} depicts the high degree of order of the DCNDBQT molecules in the film, which is promoted by a template layer consisting of 5-cyano-2-(butyl(4-phosphonic acid))-3-butylthiophene (CNBTPA). Therefore the second assumption is also fulfilled. The authors of ref. \onlinecite{Hau08} determined the molecule-molecule distance to 3.5 \AA, confirmed by a self-consistent charge density-functional based tight-binding calculation of the molecule-molecule equilibrium distance \cite{Mor09}. The third assumption of MOLED is not fulfilled because the DCNDBQT film is only a couple of monolayers thick. However, the AFM data depicts an array of well-ordered chains of $\pi$-stacked DCNDBQT molecules, hence they indicate that the transport occurs through parallel quasi-1D channels with little cross-talk.

Although this 1-dimensional approach is very straightforward it can lead to results different from the simulation of real 3-dimensional systems. MOLED assumes a homogeneous charge distribution within the molecule. Strictly speaking, this is not the case, especially close to the electrodes. Moreover, charge carriers may also be more localized at a certain region on the molecule due to polaronic effects, consequently leading to inhomogeneities of the charge distribution. Thus, future investigations are necessary to clarify this situation in our system.

For the modeling of an organic semiconductor, a detailed knowledge of its intrinsic conductance is required. In MOLED, three key mechanisms are included, which capture the physics of an organic semiconductor: First, the charge carrier tunneling from the electrodes into the channel, second the charge carrier transport via a phonon-assisted hopping process and third, charge carrier recombination and emission of photons.

\begin{figure}[ht]
        \centering
        \includegraphics[width=8cm]{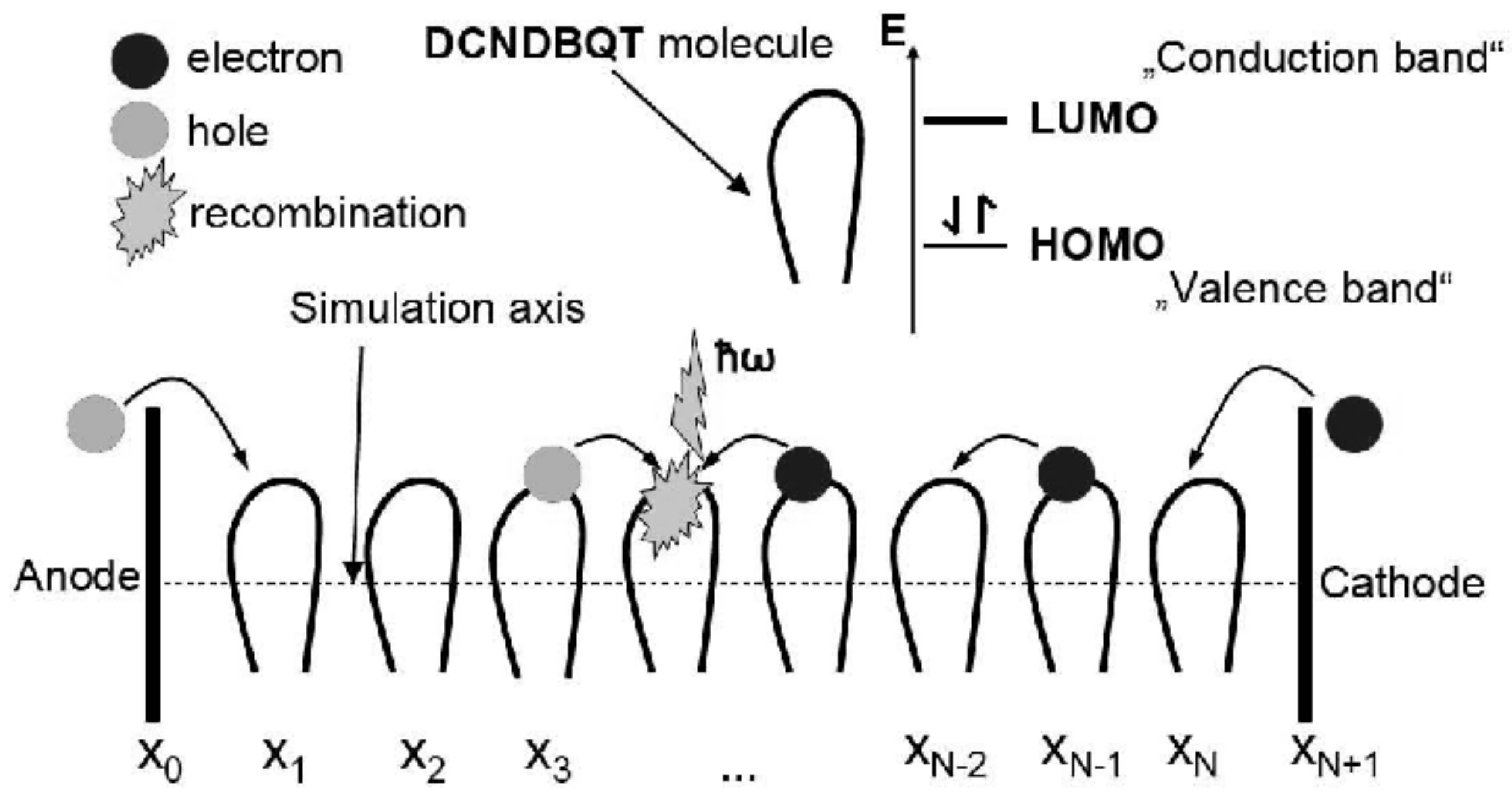}
        \caption{\label{fig:deviceSetupSketch}Sketch of MOLED setup. Molecules are situated along the simulation axis between the cathode and anode. Electrons are injected at the cathode and holes at the anode. The charge carriers move along the molecule chain via a thermally assisted hopping process. The HOMO and LUMO of the molecule correspond to the conduction and valence bands, respectively. Recombination occurs, when an electron and hole are within a characteristic distance.}
\end{figure}

Figure \ref{fig:deviceSetupSketch} gives an overview of the simulation setup and depicts these three mechanisms. The MOLED simulation axis is aligned perpendicularly to the electrodes and the hopping sites, which correspond to single molecules, are discrete in space. Thus, the molecules can be seen as flat-shaped objects on a 1-dimensional chain. Furthermore, MOLED is a two-niveau energy level approach, considering the HOMO and LUMO energy levels of the molecules for hole and electron transport, respectively.

The three main mechanisms of our simulated organic semiconductor are based on these molecular energy levels, which can be modified, for instance by Coulomb interactions, an external electric field, image forces and space-charge effects.

Consequently, the electronic states $E_m$ of an organic molecule can be characterized by:

\begin{equation} \label{eqn:energyLevel}
        E_m = \underbrace{E_{0,m}}_{reference} + \underbrace{E_{IF}(x_m) + E_{CS}(x_m) + E_D(x_m)}_{variation}
\end{equation}
with $m=1,...,N$. Equation \ref{eqn:energyLevel} determines the value of the molecular energy level of a molecule at a given position $x_m$. The first term $E_{0,m}$ is the \textit{reference term}, which corresponds to the bare energy levels of the organic molecule without any external influence factors. The HOMO and the LUMO energies are used as reference values for holes p and for electrons e, with site-specific concentrations $p_m$ and $n_m$, respectively. These bare energy levels can be determined by experimental methods \cite{Hau08} or with the help of DFT calculations \cite{Mor09}. The remaining terms of eq. \ref{eqn:energyLevel} account for the influence of the local environment of the molecule and are denoted as the \textit{variation term}. In detail, the first term $E_{IF}(x_m)$ takes into account the image force potential. The next term, $E_{CS}(x_m)$ represents the Coulomb shift of the energy levels caused by the interaction of the homogeneously charged sheets with each other. $E_D(x_m)$ is a correction term, which models an inhomogeneous charge distribution close to the electrodes instead of the general homogeneous distribution inside the device. For more details, the reader is referred to ref. \onlinecite{Hou03} and to the references therein.

\subsection{Determination of simulation-relevant parameters}

A crucial part of the model are the system-relevant OFFET parameters, for which realistic values have to be determined. Table \ref{tab:parameters} gives an overview of the employed parameter set.

\begin{table}[ht]
        \centering
        \label{Simulation parameters}
        \begin{tabular}{lc}
	\hline
	Parameter		& value \\
        \hline
        $^a$Distance between molecules & 3.5 \AA \\
        $^a$Distance molecule $\leftrightarrow$ electrode & 3.0 \AA \\
	$^a$HOMO & -5.58 eV \\
	$^a$LUMO & -3.33 eV \\
	$^c\Delta$E (HOMO) & 0.096 $\frac{eV}{GV/m}$ \\
	$^c\Delta$E (LUMO) & 0.113 $\frac{eV}{GV/m}$ \\
        $^a$Work function of electrodes (Au) & 5.1 eV \\
        $^a$DCNDBQT height & 15 \AA \\
        Discretization length $r$ & 8.74 \AA \\
        Simulation temperature & 300 K \\
        $^a$Relative permittivity of thiophene & 2.8 \\
        $^a$Hole mobility $\mu_0^p$ & 1.6 $\cdot$ $10^{-5}$ $\frac{cm^2}{Vs}$ \\
        $^{b,c}$Tunneling parameter $\gamma_{L,R}^{n,p} \cdot g_{L,R}^{n,p}$ & 0.73 electrons \\
        Tunneling prefactor $\lambda^n$ & 3.3 \AA \\
        Tunneling prefactor $\lambda^p$ & 2 \AA \\
        Attenuation factor $a^n$ & 0.7 \\
        Attenuation factor $a^p$ & 0.7 \\
        Mobility law & Poole-Frenkel \\
        $^b$Poole-Frenkel factor $F_0$ & 103416 $\frac{N}{C}$\\
        Convergence parameter & 1.0E-12 \\
        Voltage interval [$V_{min}$;$V_{max}$;$V_{steps}$] & [-3;-12;50] \\
        \hline
        \end{tabular}
        \caption{\label{tab:parameters}Overview of the employed MOLED parameters. The parameters are determined with experimental $a$, analytical $b$ or computational $c$ methods. The remaining parameters are taken from the reference model \cite{Hou03}.}
\end{table}
The parameters of category $a$ are taken from the experimental paper of ref. \onlinecite{Hau08}. Hopping is assumed between neighboring sites in reasonable agreement with the tight-binding approach of Morawetz et al. \cite{Mor09}. The underlying transport process follows the Poole-Frenkel mobility law with a calculated $F_0$ as described in ref. \onlinecite{Hou03} and \onlinecite{Fre38}. The zero-field mobilities were taken from ref \onlinecite{Hau08}. The remaining section will present a detailed discussion about the determination of the tunneling parameter. It is calculated with the help of the density of states (DOS) $g_{L,R}$ of the right and left electrode and the tunneling factor $\gamma_{L,R}$.

\begin{figure}[ht]
        \centering
        \includegraphics[width=8cm]{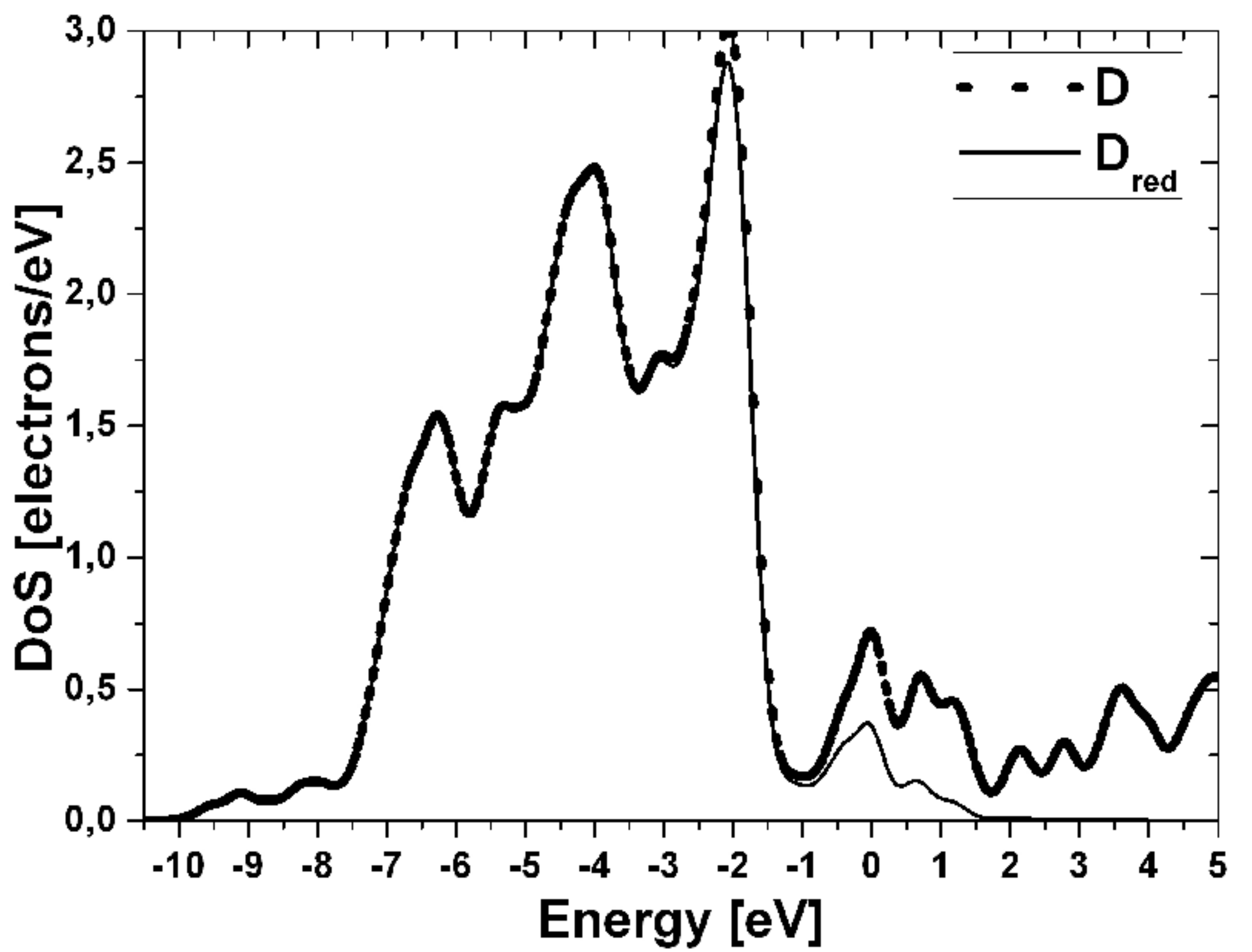} \\[5mm]
	\includegraphics[width=6.35cm]{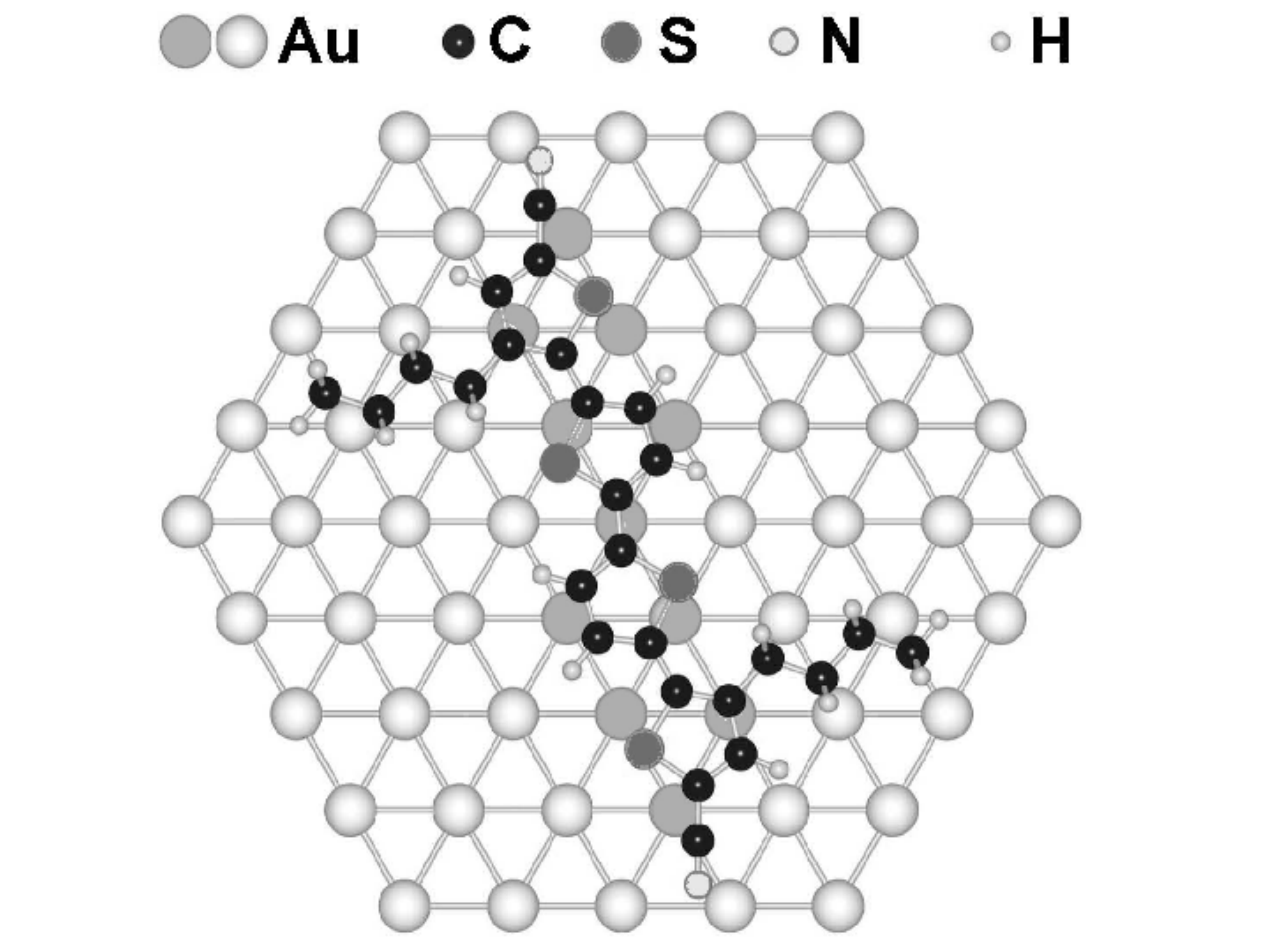}
        \caption{\label{fig:DoS_and_surface}[top] Density of states (DOS) per surface atom of an unreconstructed Au(111) surface. Energies are shifted such that zero energy corresponds to the Fermi level. The dotted curve represents the calculated DOS. The solid curve is a convolution of the calculated DOS employing the Fermi-Dirac function at a temperature of 300K. This accounts for thermally activated electrons. [bottom] Geometrical representation of the electrode - molecule interface and participating surface gold atoms to the tunneling process (highlighted grey).}
\end{figure}

The DOS is difficult to access experimentally. A suitable access to this quantity is provided in the framework of DFT, here in the plane-wave pseudopotential approach of the program package ABINIT \cite{abinit}. With such calculations it is possible to determine the DOS of the metal surface and thus estimate the contribution of mobile electrons to the transport from the electrode to the molecules at the interface. The upper panel of figure \ref{fig:DoS_and_surface} depicts the shape of the calculated DOS per gold atom of the unreconstructed Au(111) surface (dotted curve), which is the preferred surface in nanoscale structures \cite{Gem04}.

The most important contribution arises from the electrons which have an energy close to the Fermi level $E_F$. To include temperature effects one can convolute the calculated DOS per atom $D(E)$ with the Fermi-Dirac function at the expected device temperature of 300 K. The resulting curve $D_{red}(E)$ is also shown as solid line in the upper panel of figure \ref{fig:DoS_and_surface}. The value $g_{L,R}^{pA}$, which represents the fraction of electrons provided by each surface gold atom, can now be obtained by integrating the function $D_{red}(E)$ over the region modified by the Fermi-Dirac function, consequently

\begin{equation}
        g_{L,R}^{pA} = \int_{E_F}^{\infty} D_{red}(E) dE.
\end{equation}
To estimate the whole electron contribution to the tunneling, one has to determine the number of those atoms, which are accessible for the transport over the electrode-molecule surface. This can be achieved in a first order approximation by a simple geometrical estimate. The lower panel of figure \ref{fig:DoS_and_surface} depicts the geometric situation at the interface. If one considers both, the lattice parameter of gold at the surface and the area covered by the molecule, one can determine the number of gold atoms $N_{Au}$ in the vicinity of the molecule, which leads to $N_{Au} = 11$. Thus, each gold surface atom, which is situated directly below the $\pi$-orbital system of DCNDBQT, can contribute to the tunneling process. MOLED simulations with different values of $N_{Au}$ show, however, that small variations in the total number of participating gold atoms have only a minor influence on the injection term, which indicates, that the model of the electrode-molecule interface is insensitive against small geometric variations. The total value of $g_{L,R}$ can be determined by

\begin{equation}
        g_{L,R} = N_{Au} g_{L,R}^{pA}.
\end{equation}
The second parameter $\gamma_{L,R}$ is also difficult to determine experimentally, because the employed molecule DCNDBQT has a negligible tunneling resistance \cite{Eng09}. Instead, a value can be deduced from an analytical solution, where the Schottky barrier is calculated for a mean source-drain field of $\bar{F}=0.65 \frac{MV}{cm}$.

Here, the electrode is considered as a confining potential well, which restricts the motion of the mobile electrons. During the injection process, the electron has to overcome a characteristic potential $V$ to tunnel directly from the metal to the molecule. In this approach, the transmission coefficient $T$, which denotes the probability of the electron transmission, can be calculated analytically as

\begin{equation}
        T = \Biggl[1 + \frac{V^2}{4E(V-E)} \sinh^2(\kappa)\Biggr]^{-1}
\end{equation}
where $E$ is the energy of the electron wave, $V$ the potential barrier between electrode and molecule, and

\begin{equation}
        \kappa = \frac{b \sqrt{2m(V-E)}}{\hbar}.
\end{equation}
$m$ is the mass of the electron and $b$ the tunneling distance. From the assumption that $\kappa$ fulfills $\kappa \gg 1$, it follows for the transmission coefficient $T$

\begin{equation}
        T \approx \frac{16E(V-E)}{V^2} \exp(-2 \kappa).
\end{equation}
The prefactor of the exponential term can be identified with the simulation parameter $\gamma_{L,R}$ (see for instance eq. 5 of ref \onlinecite{Hou03}), consequently

\begin{equation}
        \gamma_{L,R} = \frac{16E(V-E)}{V^2}
\end{equation}
The remaining task is to determine the potential barrier $V$, which is a function of the undistorted barrier $\Phi_B$ of the molecule-electrode interface, the image force (second term) and the field contribution (third term in eq \ref{analytic}). Consequently, 

\begin{equation} \label{analytic}
	V = \int_{0}^{b} (\Phi_B - \frac{q^2}{16\pi\epsilon_0\epsilon_rx} - q\bar{F}x) dx
\end{equation}
had to be integrated over the electrode - molecule distance of $b=3$ \AA. This electrode - molecule distance is in agreement with literature data \cite{Min08}. In the physical picture, the electron energy $E$ for tunneling through the barrier can be approximated by the Fermi energy $E_F$ of the metal electrode.

\section{Results}
\label{Results}

In this section we present the results on the OFFET performance for a device with an electrode-electrode distance of 1000 {\AA} at different gate voltages of $V_g=\{0;1;2\}$ volts. These voltage values represent an effective polarization field of $F_{tr}=\{0;0.5;1\}\frac{GV}{m}$ in our 1D thiophene stack in accordance to the experimental electric fields \cite{Hau08}.

The device distance was chosen to model a realistic extend of the ferroelectric domain of the gate \cite{Eng09}. Thus, the switching of the ferroelectric substrate will polarize the whole area between the leads. We will investigate quantitatively the influence of the polarization on the charge carrier density at constant source-drain field. 

\begin{figure}[ht]
        \centering
	\includegraphics[width=8cm]{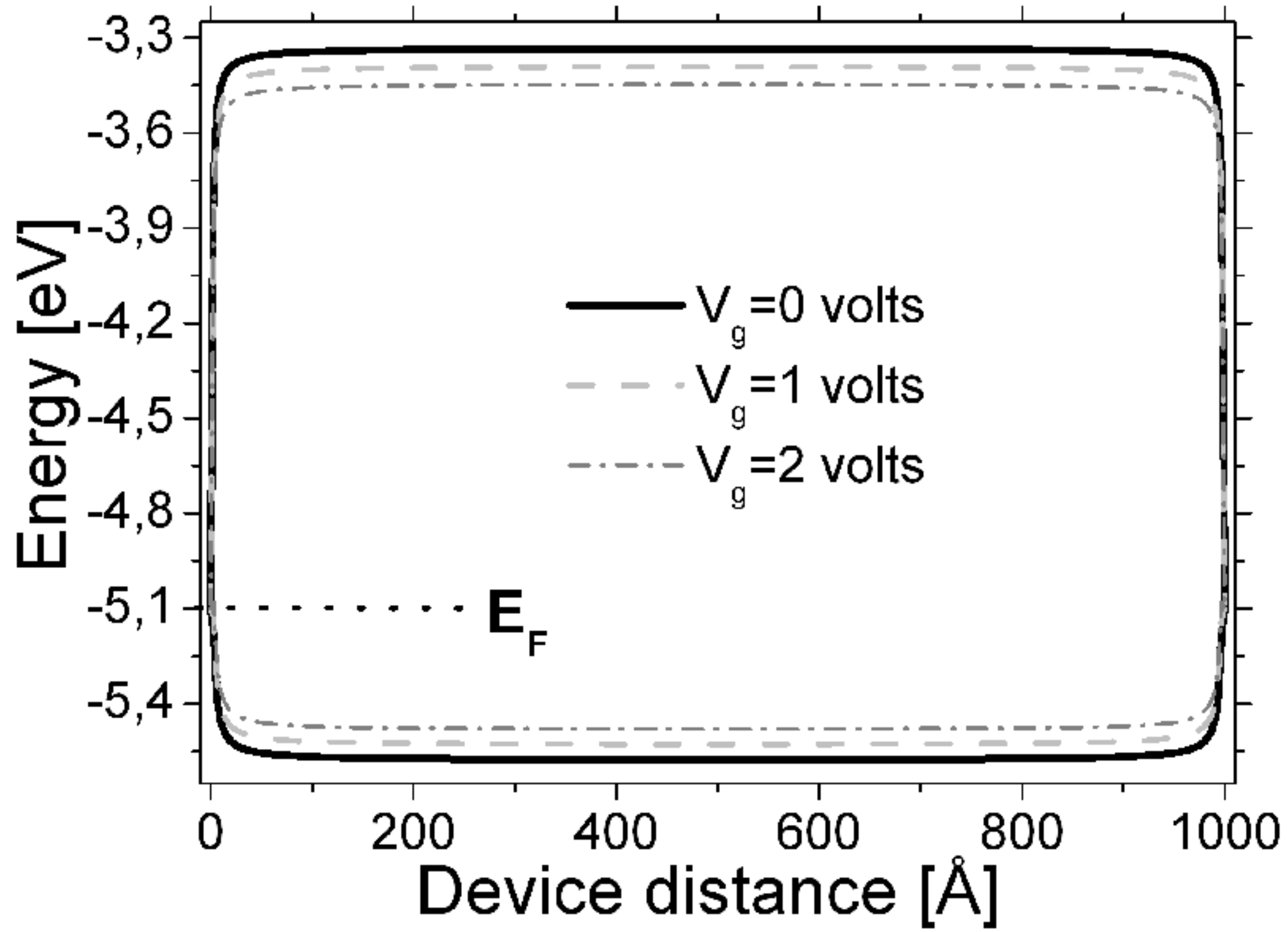} \\[5mm]
	\includegraphics[width=8cm]{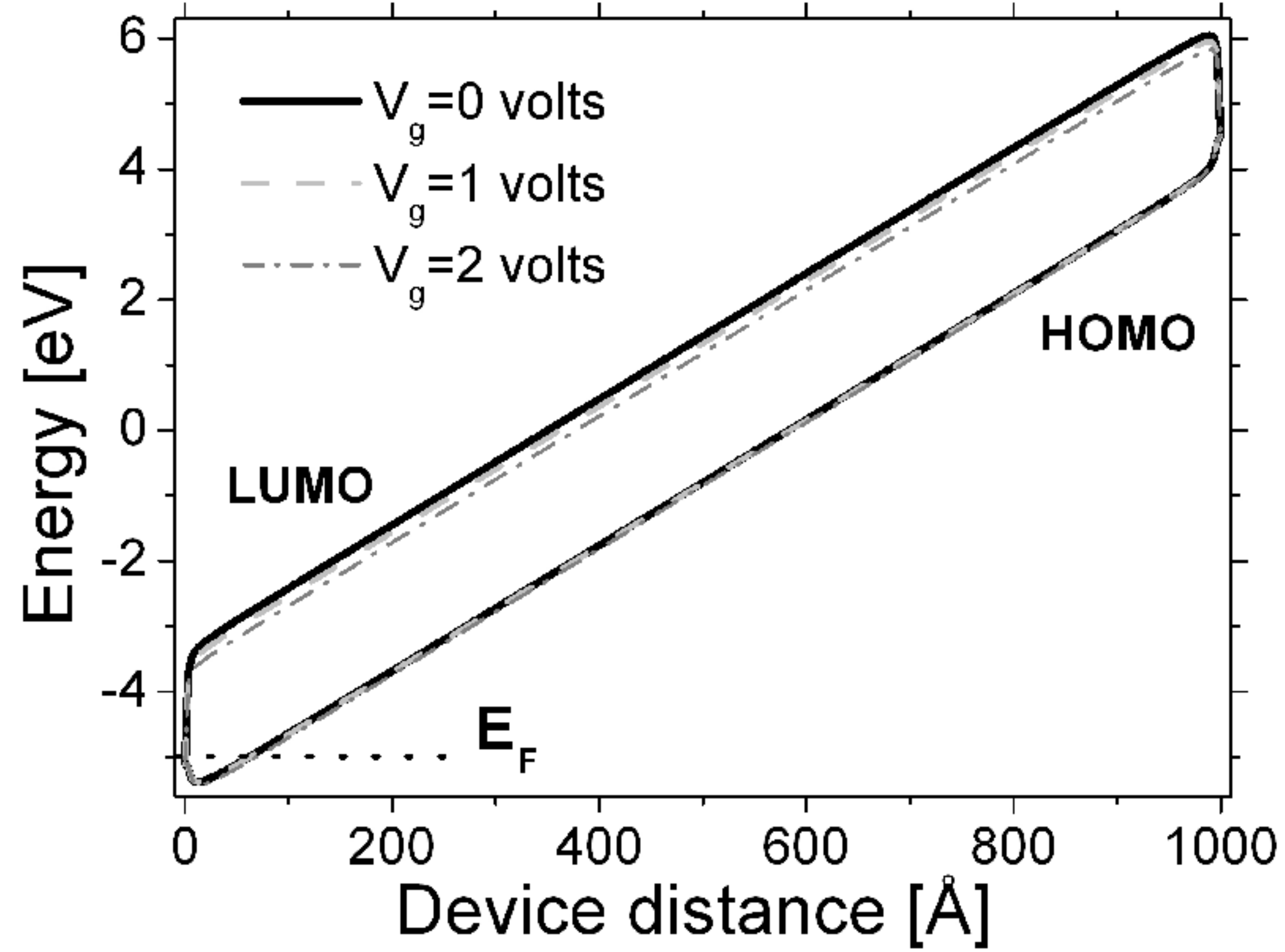}
        \caption{\label{fig:eva1}HOMO and LUMO energy levels for different gate voltages $V_g$ in absence (upper panel) and presence (lower panel) of an applied source-drain voltage of $V_{SD}\approx 10$ volts. The Fermi level of the gold electrode is also shown.}
\end{figure}

Figure \ref{fig:eva1} depicts the HOMO and LUMO energy levels without (upper panel) and with (lower panel) an applied source-drain voltage of $V_{SD}\approx 10$ volts.

Although the energy levels are localized molecular states and do not form a band, a continous representation was chosen, because the device is much larger than the intermolecular spacing. Both panels of figure \ref{fig:eva1} shows also the position of the Fermi level of the gold electrodes. Obviously, the injection barrier $E_B$ for holes, which enters from the left side into the device, is much lower in comparison to the electron injection barrier on the cathode side. We estimated $E_B$ for holes as $E_B^p \approx 0.3$ eV and for electrons as $E_B^e \approx 1.8$ eV, respectively. Consequently, the holes can be considered as the majority charge carrier class, in accordance with experiment \cite{Hau08}.

\begin{figure}[ht]
        \centering
	\includegraphics[width=8cm]{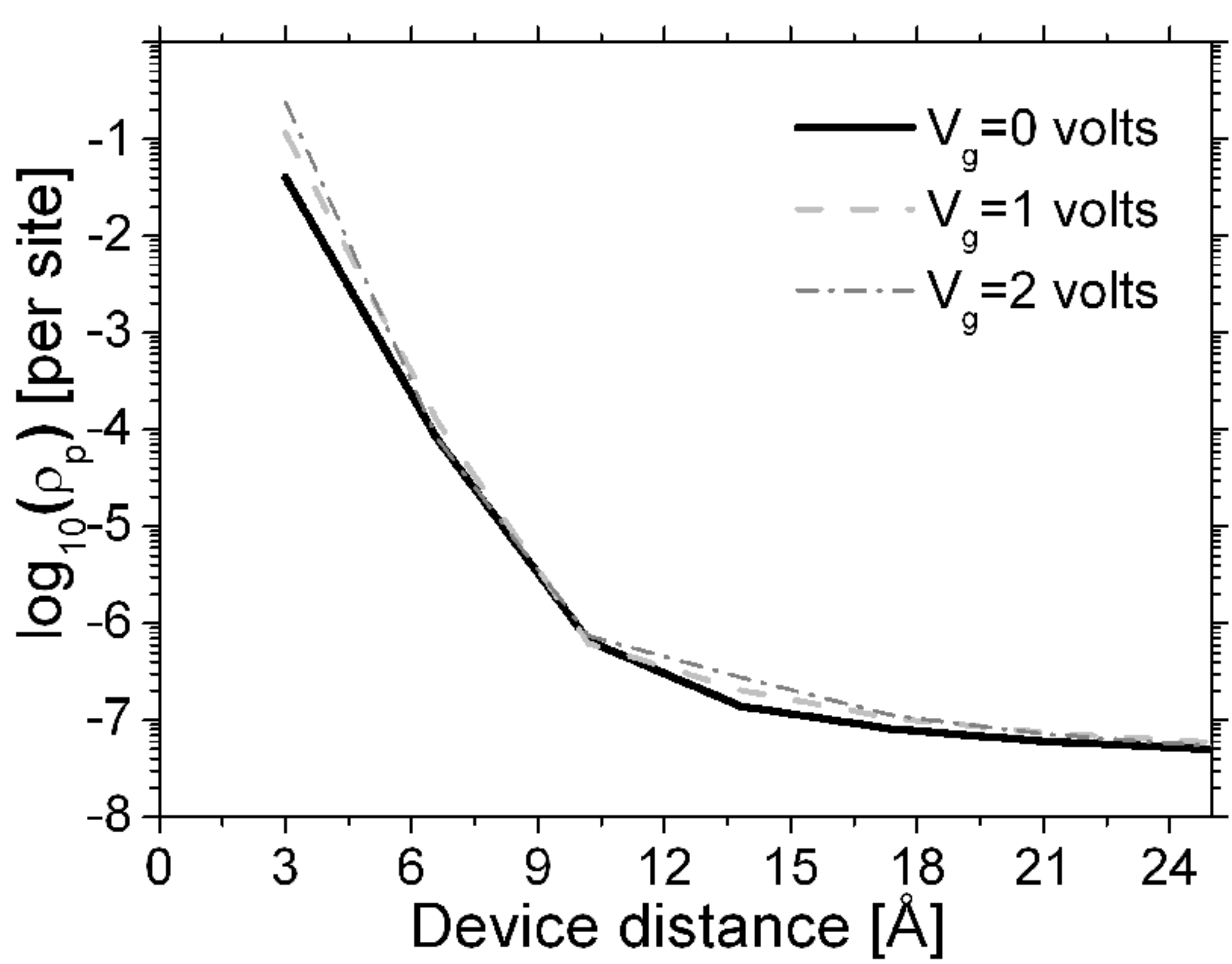} \\[5mm]
	\includegraphics[width=8cm]{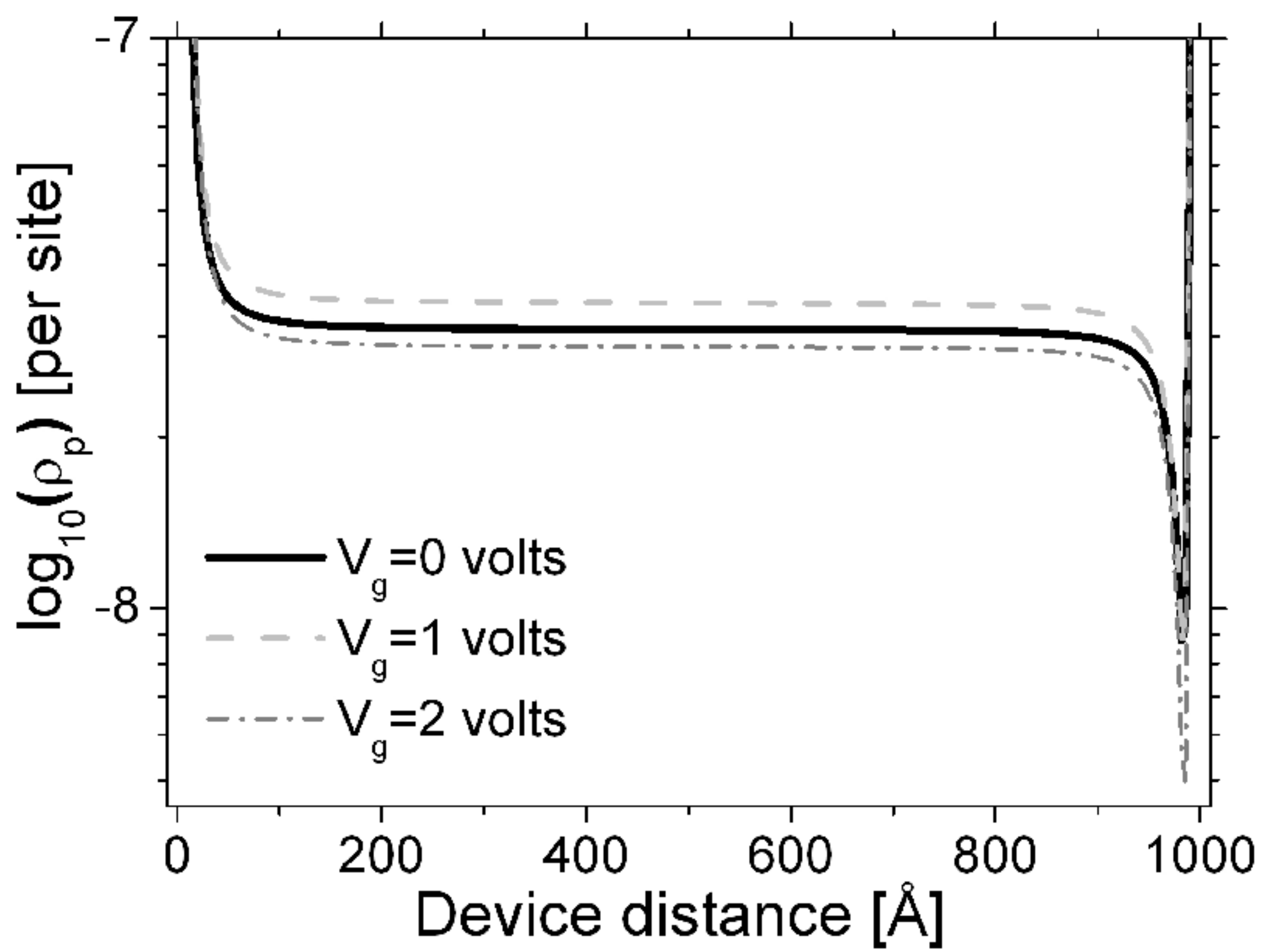}
        \caption{\label{fig:eva2}Hole density (majority charge carrier class) depending on the device position. [Top] A magnification of the anode-organic interface. [Bottom] The hole density of the whole device.}
\end{figure}

An applied gate voltage $V_g$ will decrease the HOMO-LUMO energy gap, as indicated in the upper panel of figure \ref{fig:eva1}. We expect, that there might occur an increase of the hole carrier density inside the device due to the lowering of the energy barrier for injection. Indeed, a closer look at the first molecules near the anode confirm an increase of the hole carrier density with increasing gate voltage $V_g$, as depicted in the upper right panel of figure \ref{fig:eva2}.

\begin{figure}[ht]
        \centering
        \includegraphics[width=8cm]{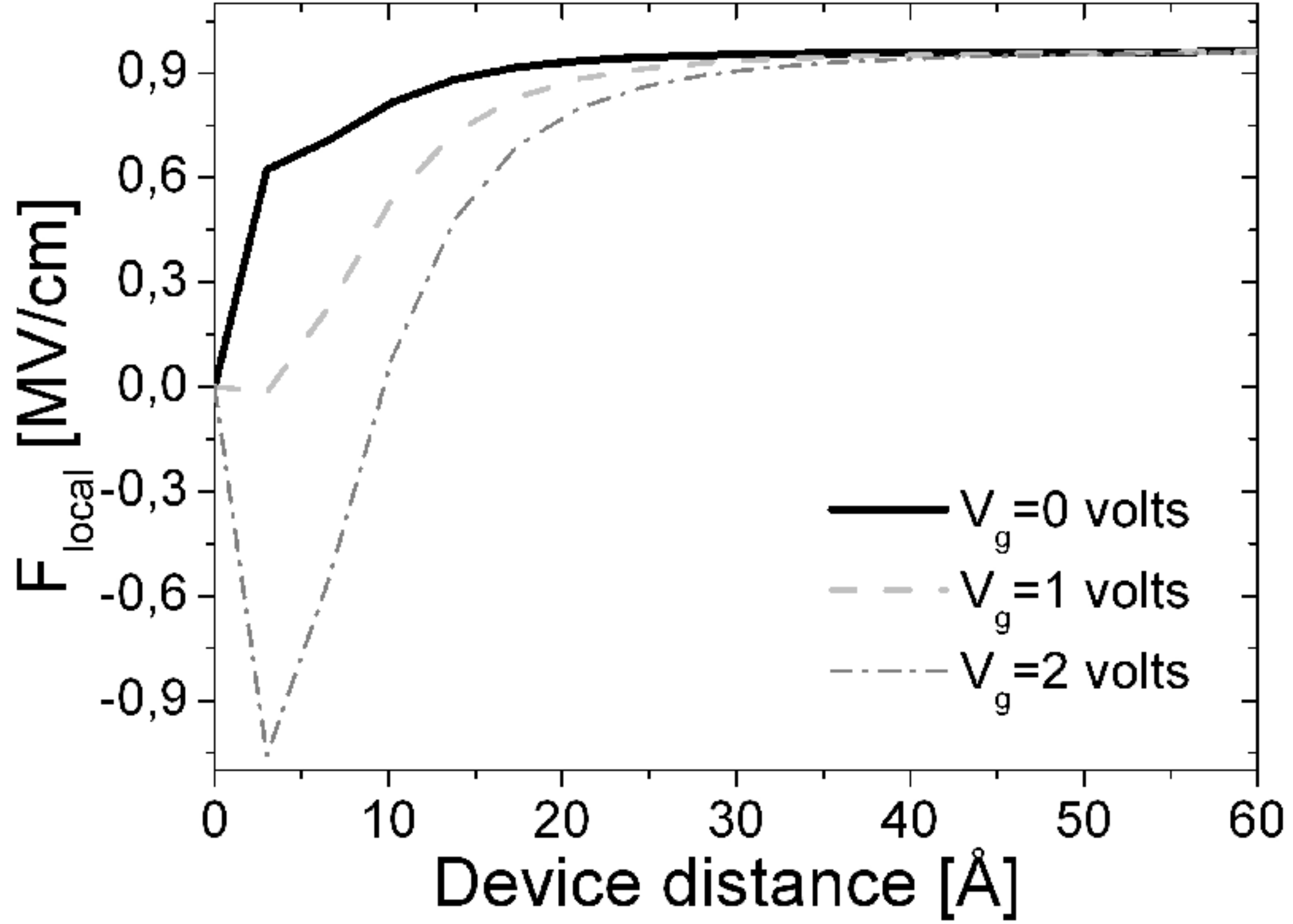}
        \caption{\label{fig:eva3}[left] Local electric field in dependence of the applied gate voltage.}
\end{figure}

However, the lower panel of figure \ref{fig:eva2} shows a decrease of the hole density further inside the device for a gate voltage of $V_g=2$ volts. This can be explained by the image force contribution, which arises at the metal-organic interface. The accumulated holes at the interface introduce image charges within the electrode, which act as an effective Coulomb blockade. Consequently, a counteracting field against the source-drain bias is created, which is shown in figure \ref{fig:eva3}. This effect plays a minor role for gate voltages below or equal to $V_g=1$ volt.

Additionally, this effect becomes pronounced due to the low hole carrier mobility of $\mu_0\sim10^{-5}\frac{cm^2}{Vs}$, by which the charge carriers are not transported away sufficiently fast from the injecting electrode.

\begin{figure}[ht]
        \centering
	\includegraphics[width=8cm]{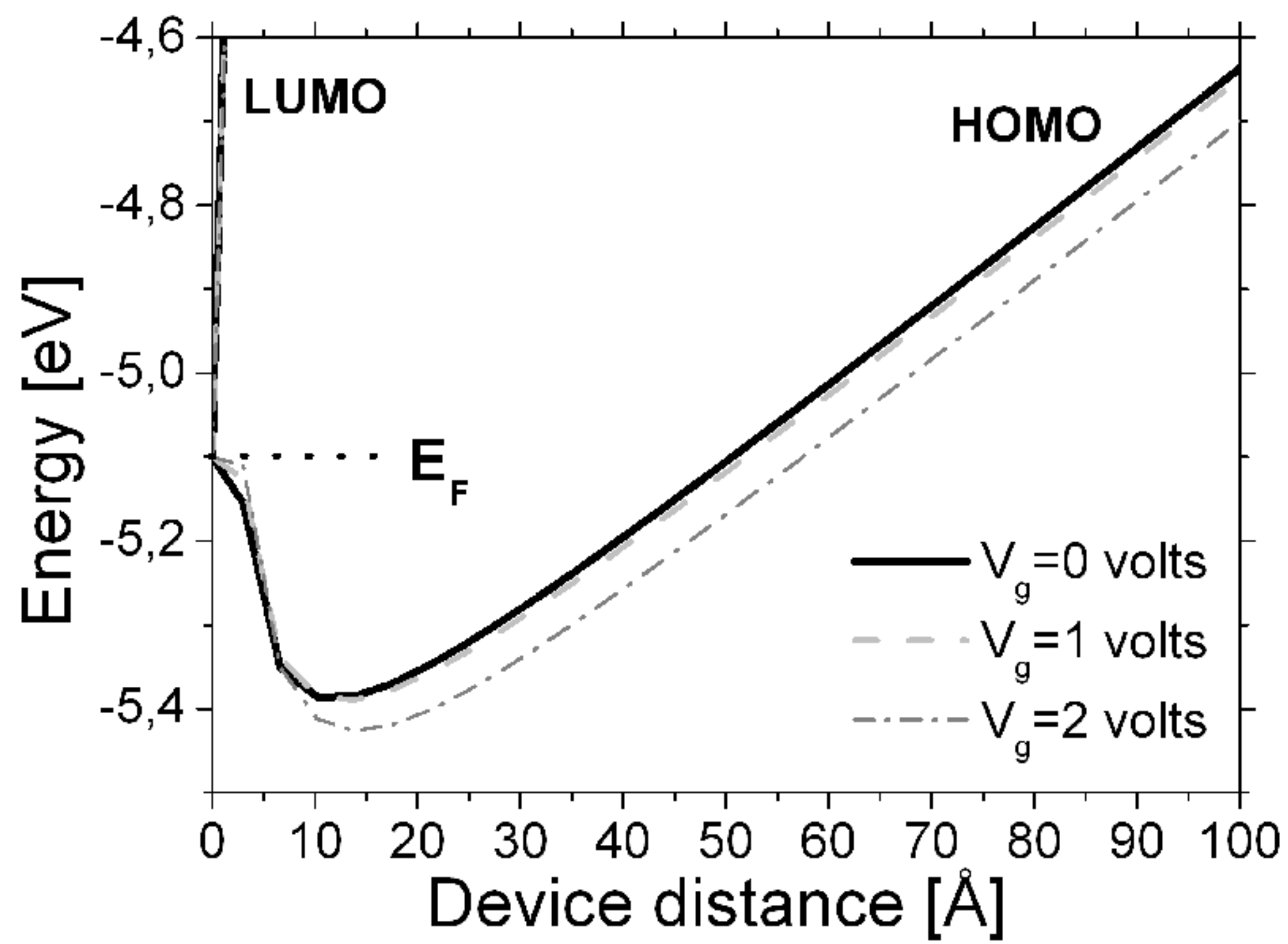}
        \caption{\label{fig:eva4}[right] Variation of the Schottky barrier due to the change of the image force contribution.}
\end{figure}

Figure \ref{fig:eva4} depicts the HOMO energy levels of the first 26 molecules near the anode-organic interface. Obviously, the change in the image force contribution influences the shape and position of the Schottky barrier. This Schottky barrier is slightly elevated, which increases the effective energy barrier for charge carrier injection, consequently leading to a decrease of the hole carrier density further inside the device.

\section{Summary \& Outlook}
\label{Conclusion}

The electronic transport of electrons and holes through stacks of $\alpha$,$\omega$-dicyano-$\beta$,$\beta$'-dibutyl- quaterthiophene as part of a novel organic ferroic field-effect transistor was presented.

We investigated the influence of the ferroelectric polarization of the substrate on the electronic transport through the organic layer with the help of the phenomenological model MOLED, which treats the hopping transport of electrons and holes in such films by a master equation approach. To accomplish that, an extension of MOLED to include the influence of transverse electric fields was essential. The effect of the transverse field can be easily modeled by a linear decrease of the HOMO-LUMO gap for moderate field strengths, which we have calculated using the SCC-DFTB method.

A subtask of this study was the determination of the simulation-relevant parameters by experimental, analytical and computational methods. We presented detailed approaches for the determination of two essential simulation parameters, the density of states at the gold electrode and the tunneling prefactor for injection into the organic medium. Both parameters are hardly accessible by experimental methods for this special kind of device.

After determining the MOLED parameters, we qualitatively studied the performance of a fully polarized OFFET device. We investigated the device behavior for different gate voltages and showed the capability of the model for organic device analysis.

Further investigations will focus on the comparison with prototypical devices and a detailed understanding of the underlying physics.

\acknowledgements

The authors thank the Deutsche Forschungsgemeinschaft (`Integrierte elektrokeramische Funktionsstrukturen` DFG-SPP1157) for funding. The financial support by the Brazilian Ministry of Science and Technology is acknowledged.

\bibliography{ofet}

\end{document}